\documentclass[journal=nalefd,manuscript=letter]{achemso}
\usepackage[version=3]{mhchem} 
\usepackage{color}
\usepackage{soul}
\usepackage{graphicx}
\usepackage{epstopdf}
\usepackage{upgreek}
\usepackage{amsmath}
\usepackage{amssymb}
\usepackage{amsfonts}

\author{Erik M\r arsell}
\altaffiliation{Current address: University of British Columbia, Quantum Matter Institute, 2355 East Mall, Vancouver, BC  V6T 1Z4, Canada}
\author{Emil Bostr\"om}
\author{Anne Harth}
\author{Arthur Losquin}
\author{Chen Guo}
\author{Yu-Chen Cheng}
\author{Eleonora Lorek}
\author{Sebastian Lehmann}
\author{Gustav Nylund}
\author{Martin Stankovski}
\author{Cord~L. Arnold}
\author{Miguel Miranda}
\author{Kimberly~A. Dick}
\author{Johan Mauritsson}
\author{Claudio Verdozzi}
\author{Anne L'Huillier}
\author{Anders Mikkelsen}
\email{anders.mikkelsen@sljus.lu.se}
\affiliation{Lund University, Department of Physics, PO Box 118, 221 00 Lund, Sweden}

\title{Spatial Control of Multiphoton Electron Excitations in InAs Nanowires by Varying Crystal Phase and Light Polarization}

\abbreviations{IR,NW,GW,PEEM}
\keywords{Multiphoton photoemission, Polytypism, III--V, Semiconductor nanowires, Nonlinear optics, Band structure}

\begin{document}

\begin{tocentry}

\includegraphics[width = 9cm,height=3.5cm]{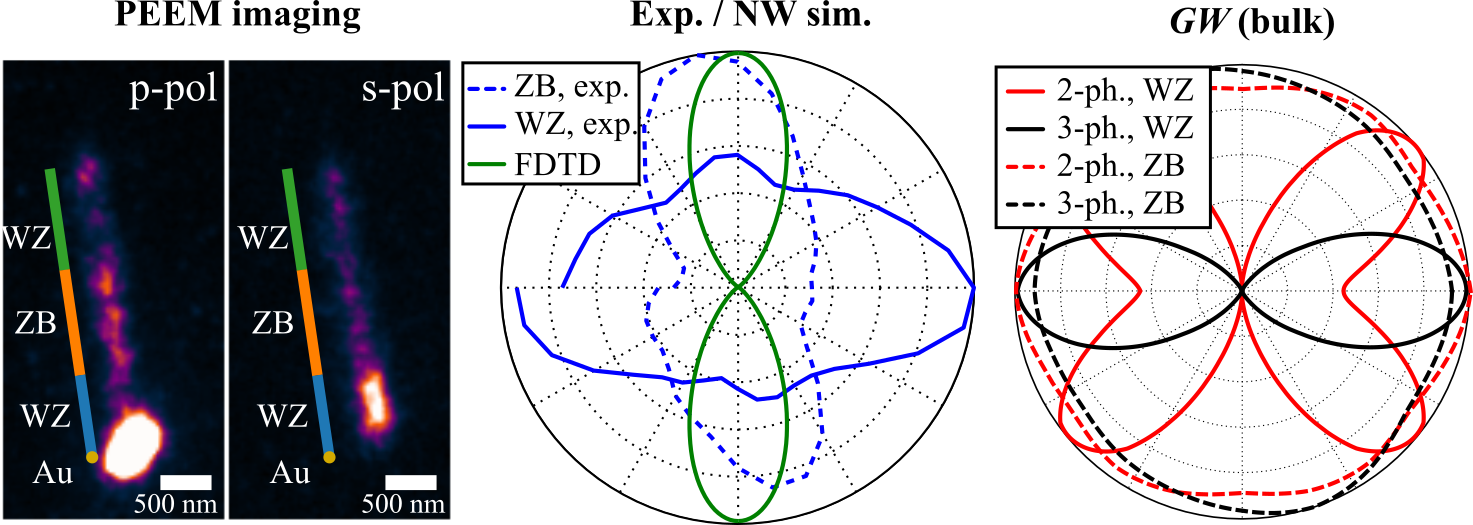}
\end{tocentry}

\begin{abstract}
We demonstrate the control of multiphoton electron excitations in InAs nanowires (NWs) by altering the crystal structure and the light polarization. Using few-cycle, near-infrared laser pulses from an optical parametric chirped-pulse amplification system, we induce multiphoton electron excitations in InAs nanowires with controlled wurtzite (WZ) and zincblende (ZB) segments. With a photoemission electron microscope, we show that we can selectively induce multiphoton electron emission from WZ or ZB segments of the same wire by varying the light polarization. Developing \textit{ab-initio GW} calculations of 1st to 3rd order multiphoton excitations and using finite-difference time-domain simulations, we explain the experimental findings: While the electric-field enhancement due to the semiconductor/vacuum interface has a similar effect for all NW segments, the 2nd and 3rd order multiphoton transitions in the band structure of WZ InAs are highly anisotropic, in contrast to ZB InAs. As the crystal phase of NWs can be precisely and reliably tailored, our findings opens up for new semiconductor optoelectronics with controllable nanoscale emission of electrons through vacuum or dielectric barriers.
\end{abstract}

When matter is subjected to sufficiently large electric fields, nonlinear processes such as electron excitation by simultaneous absorption of multiple photons can strongly influence the optical response. Through this multiphoton electron excitation process, highly energetic electrons can be produced from visible or near-infrared light. These electrons can then either lead to cascaded excitations of multiple electrons inside the material, or leave the material into bordering vacuum or another material (electron emission). Such energetic electrons are relevant for a many different applications such as hot-carrier photovoltaics~\cite{Kolodinski93,Dong16}, new types of electronics based on photoemission~\cite{Forati16}, lasers~\cite{He16}, and electron emitters~\cite{Tseng05,Lin05,Swanwick14}. 

The strong electric fields needed for efficient multiphoton electron excitation can be reached by concentration of light in time and/or space. Concentration of light in space can be done using nanostructured surfaces, while concentration in time is achieved by using ultrashort laser pulses. The rapid development of both nanofabrication methods and ultrafast laser sources has strongly increased the importance of nonlinear effects in solids. In particular, by concentrating light it is possible to sustain high peak field strengths without damaging the material~\cite{Gertsvolf08}, and nonlinear effects such as multiphoton electron excitation can become important even at moderate excitation field strengths~\cite{Aeschlimann95}. However, imaging of nonlinear effects in semiconductors or dielectrics with subwavelength resolution remains a largely unexplored field.

Multiphoton photoemission electron microscopy (PEEM), is a widespread technique for characterizing multiphoton electron excitations at nanostructure surfaces with subwavelength resolution~\cite{Schmidt01,Kubo05,Aeschlimann07}. In contrast to many other methods it has the benefit of combining the control of temporal and spectral properties of the exciting optical fields with a spatial resolution only limited by electron optics, and possibly reaching the few-nm regime~\cite{Tromp10}. Multiphoton PEEM has been employed in ultrafast nano-optics mainly using noble metal nanostructures because of their surface plasmon resonances facilitating nanoscale concentration of optical energy in the visible part of the spectrum. In such studies, strong polarization dependences due to the nanostructure morphology have been used for spatial control of the field localization, and thus the multiphoton electron excitations~\cite{Aeschlimann07,Word11,Marsell15}. 

The multiphoton photoemission from metal nanostructures can in most cases be well described classically by a coherent, isotropic, plasmon-assisted process whose efficiency qualitatively only depends on the shape of the object and the linear, isotropic dielectric function~\cite{Merschdorf04,Aeschlimann10}. This description relies on several assumptions, including an isotropic linear and nonlinear microscopic material response, as well as a negligible heating of the electron gas. However, for lower-symmetry materials exhibiting a non-cubic crystal structure, such as the hexagonal wurtzite (WZ) structure, the assumption of an isotropic microscopic response might not be valid. The multiphoton electron excitation efficiency can in this case be expected to depend on a combination of at least three geometrical parameters: the light polarization, the surface morphology, and the crystallographic orientation. In most cases, the effects of polarization and morphology can be well modeled using Maxwell's equations and tabulated values of the linear dielectric function. However, predicting the effect of crystallographic orientation on the multiphoton transitions presents new computational challenges as it requires reliable calculations of the electronic states and transition matrix elements for bands high above the Fermi level. This can be achieved using \textit{GW} calculations, which are, however, challenging to perform over a dense enough grid of k-points.

III--V semiconductor nanowires (NWs) are a relevant model system for studying how morphology and crystal structure affects the material properties, as both of these parameters can be tailored with high precision. Furthermore, they show great promise for a variety of applications within e.g. photovoltaics~\cite{Wallentin13}, lighting~\cite{Svensson08}, and electronic device technology~\cite{Egard10,Riel14}. In particular, the crystal structure is an interesting characteristic of these NWs, as they, in contrast to their bulk counterparts, can be controllably grown with segments of both the cubic zincblende (ZB) and the hexagonal wurtzite (WZ) phase. This fundamental difference in crystal symmetry can affect the response to any physical probe with directionality, be it optical, electrical, or mechanical. Indeed, the crystal structure can affect, for example, the morphology~\cite{Knutsson15}, electronic properties~\cite{Hjort14}, transport properties~\cite{Thelander11,Ullah13}, and the linear optical response~\cite{Spirkoska09,Anttu14} of the materials. However, very few studies exist on the influence of crystal symmetry on the nonlinear interaction of light with semiconductors, as this poses both experimental and theoretical challenges.

In this study, we show how multiphoton electron excitations in semiconductors can be controlled by the polarization of the light and the crystal structure of the material. We do this by imaging the polarization-dependent multiphoton electron emission from InAs NWs with alternating WZ and ZB segments. Our experimental approach is based on the combination of PEEM with broadband few-cycle infrared laser pulses from a non-collinear optical parametric amplifier. This adds to the few recent multiphoton PEEM studies on semiconductor surfaces~\cite{Fitzgerald13,Fukumoto14,Aeschlimann15,Man16}, and is the first on semiconductor NWs. In a theoretical approach, we develop the theoretical machinery and calculate polarization-dependent multiphoton transition rates from self-consistent quasi-particle \textit{GW} band structure calculations~\cite{Hedin65,Godby86,Aryasetiawan98}. This is used for evaluation of the influence of the microscopic effects of the crystal symmetry. The influence of the nanostructure morphology is investigated based on classical electrodynamics using the finite-difference time-domain (FDTD) method. The experimentally observed multiphoton photoemission is seen to be strongly polarization dependent, with different polarizations favored for different crystal structures. For ZB InAs, the experimental results can be understood by only considering the nanostructure morphology. For WZ InAs however, strongly anisotropic multiphoton transitions dominate the polarization dependence of the photoemission signal, as confirmed by the microscopic \textit{GW} calculations. In this system, we can therefore selectively control multiphoton electron excitations by adjusting the material crystal structure and the polarization of the excitation light.

\begin{figure}%
\includegraphics[width=3.33 in]{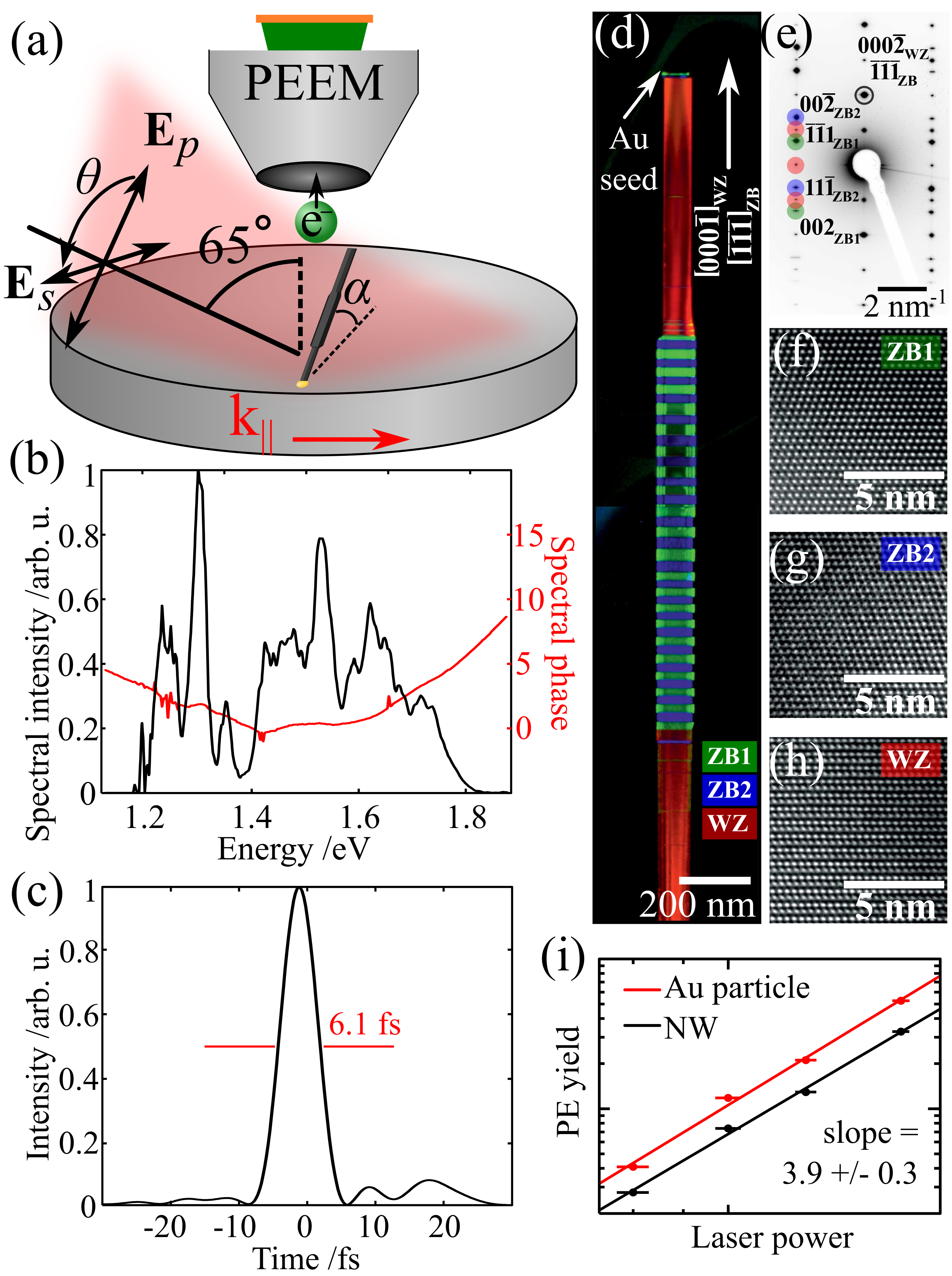}%
\caption{Description of the experiment. (a) Schematic drawing of the setup. Femtosecond laser pulses impinge at a grazing incidence and with varying polarization characterized by the angle $\theta$, which is the clockwise deviation from $p$-polarization. The nanowires are lying on the substrate with an angle $\alpha$ to the vertical direction, which is perpendicular to the plane of incidence. (b) Spectral intensity and phase of the laser pulse. (c) Temporal intensity profile of the laser pulse, showing a full-width at half-maximum duration of 6.1~fs. (d) Convoluted conventional transmission electron microscopy dark field image of a NW, indicating the different crystal phases and orientations WZ (red), ZB twin 1 (green), ZB twin 2 (blue). The Au seed particle, appearing dark in this imaging mode, is located at the top. Some of the crystal phase specific diffraction spots are highlighted in the selected area diffraction pattern in (e), including the ones chosen for the crystal phase selective imaging in (d). (f--h) High resolution TEM images of the different crystal phases acquired from a NW from a different sample but grown on an identical substrate and under identical growth conditions. (i) Double-logarithmic plot of the photoelectron intensity as a function of laser power, indicating a 4-photon photoemission process for both Au and for the NW.}
\label{fig:Schematic}%
\end{figure}

\paragraph{Experiments.} The experimental setup is illustrated in Fig.~\ref{fig:Schematic}a and is based on a commercial PEEM instrument (Focus GmbH) and a two-stage optical parametric amplifier system (OPCPA) operating at 200~kHz repetition rate. Laser pulses of 6.1~fs duration, 850~nm center wavelength and $\sim 10^{11}$~W~cm$^{-2}$ peak intensity impinged on the sample at a 65$^\circ$ angle from the sample normal. The linear polarization was varied using a broadband half wave plate. The spectrum and phase of the laser pulses (Fig.~\ref{fig:Schematic}b--c) were characterized using the dispersion scan (d-scan) technique~\cite{Miranda12}. The short pulse duration favors photoemission by direct transitions in the material, rather than emission from a thermalized hot electron distribution as the electron thermalization time is typically longer than the laser pulse duration~\cite{Suemoto98}. To confirm that the photoemission signal is not dominated by excitations with lifetimes longer than the pulse duration, we also performed pump--probe measurements (see Supporting Information) using two laser pulses, as previously done on metallic samples~\cite{Marsell15NL}. The measured delay-dependent photoemission intensity from the semiconductor NWs followed an envelope similar to that from the Au metal particle at the end of the NW, consistent with decay processes on a time scale on the order of the laser pulse duration or shorter.

Nanowires (NWs) were grown using Au seed particles defined by electron beam lithography and transferred to a highly doped Si substrate for PEEM imaging. A convoluted dark-field transmission electron microscopy (TEM) image of a typical NW (Fig.~\ref{fig:Schematic}d--e) shows the high crystal phase purity and the abrupt interfaces between the WZ and ZB segments. The growth direction is indicated with an arrow. High-resolution TEM images from a sample grown under identical growth conditions confirm the purity of the crystal structures (Fig.~\ref{fig:Schematic}f--h). As the growth method used for crystal phase switching relies on changing the III--V ratio~\cite{Lehmann13}, no dopants have been purposely introduced in any of the segments. The Au particles could in principle be removed via preferential etching. However, keeping the Au particles on the NWs ensures a smooth NW surface, minimizes possible contaminants, and provides a reference metallic particle for the multiphoton PEEM experiments.

The central energy of the laser pulses of 1.5~eV indicates that multiple photons must be absorbed for each emitted electron. Assuming a work function of Au of 5.1~eV~\cite{Michaelson77} and an electron affinity of InAs of 4.9~eV~\cite{Milnes72}, we expect a 4-photon process for both of these materials. To verify that we were in the multiphoton photoemission regime, we measured the photoemission yield as a function of laser power (see Fig.~\ref{fig:Schematic}i). The power law dependence was found to be the same both for the NW and the Au particle and for the different parts of the NW, within the experimental precision. The observed power of 4 dependence agrees well with a perturbative multiphoton photoemission process for both Au and InAs.

\begin{figure}%
\includegraphics[width=3.33 in]{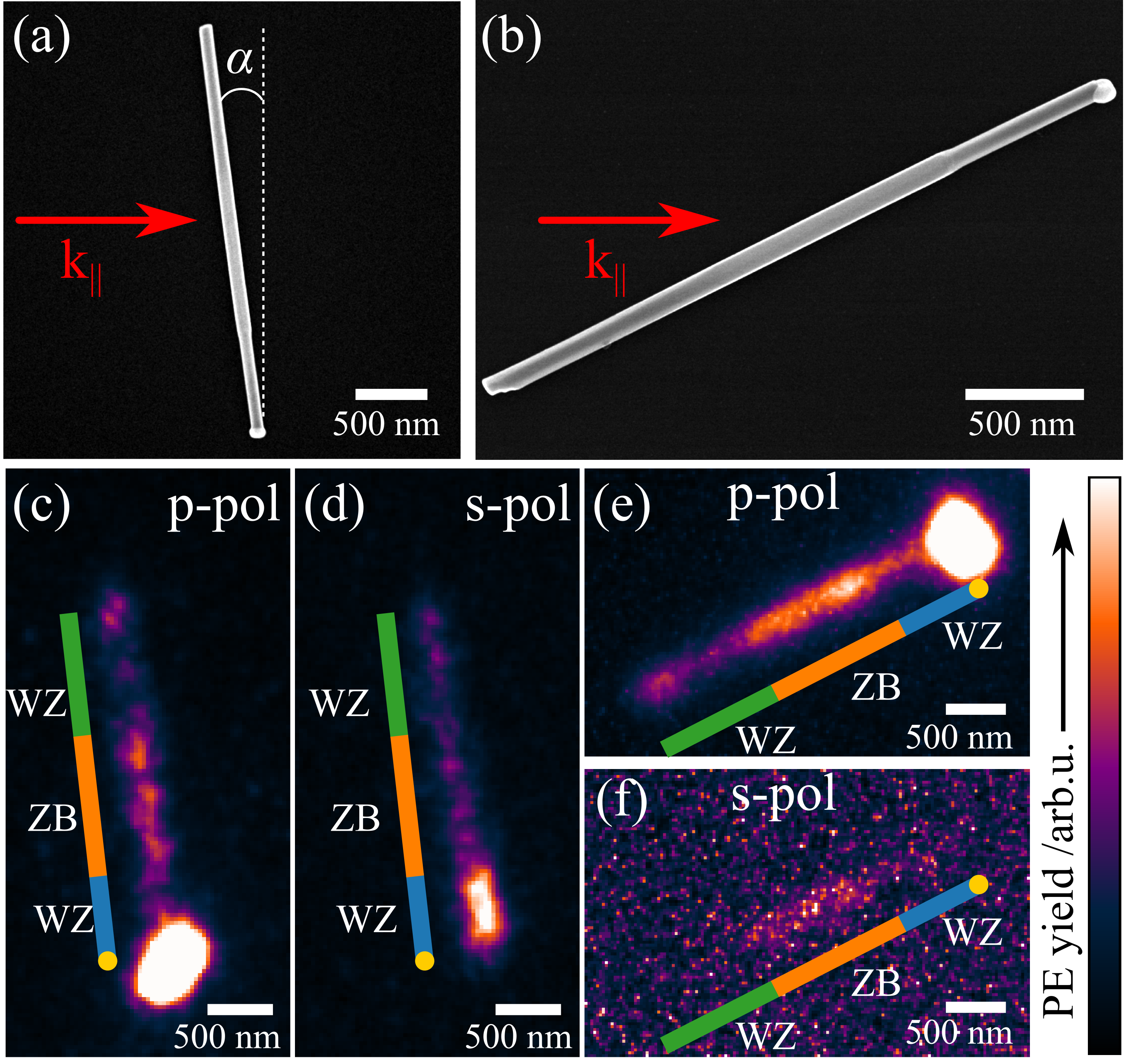}%
\caption{Demonstration of different spatial distributions of emitted photoelectrons.(a--b) SEM images of two example nanowires with different orientations with respect to the plane of incidence, $\alpha = 7^\circ$ (a) and $\alpha = 117^\circ$ (b). The in-plane direction of the incident light is indicated with red arrows. (c--f) Multiphoton PEEM images of the NWs depicted in (a--b) displaying (c) Preferential emission from ZB middle segment (polarization is $p$, wire from (a)) (d) Preferential emission from ``top" WZ segment, closest to the Au particle (polarization is $s$, wire from (a))  (e) Preferential emission from the ZB middle segment (polarization is $p$, wire from (b)) (f) Quenched emission from all NW segments (polarization is $s$, wire from (b)). In all PEEM images, a model of the NW is shown as an inset to visualize the emission from the different segments.}
\label{fig:PEEMimgs}
\end{figure}

To demonstrate the ability to selectively induce multiphoton electron excitations in different parts of the nanowires, we show typical multiphoton PEEM images in Fig.~\ref{fig:PEEMimgs}c--f. The two nanowires have similar lengths and morphologies, but are rotated differently with respect to the plane of incidence of the beam (see SEM images in Fig.~\ref{fig:PEEMimgs}a--b). The two NWs can be characterized by the angle of rotation $\alpha$, defined in Figs.~\ref{fig:Schematic}a and \ref{fig:PEEMimgs}a, leading to $\alpha = 7^\circ$ (a) and $\alpha = 117^\circ$ (b), respectively. For both NWs, the emission from the Au particle is maximized for $p$-polarization, reaching higher values than the emission from the NW (Fig.~\ref{fig:PEEMimgs}c,e). 
The ZB (middle) segment of the nanowires also shows a stronger emission than the WZ segments for $p$-polarized excitation. However, when excited by $s$-polarized light (Fig.~\ref{fig:PEEMimgs}d,f), the two nanowires show very different photoemission behavior. While the NW in (b) shows almost no detectable emission, the signal from the NW in (a) is dominated by the upper WZ segment (closer to the Au particle). The emission from the lower WZ segment is comparable to that from the ZB segment for this polarization. This change in image contrast between the ZB segment and the upper WZ segment when switching from $s$- to $p$-polarization on NWs almost perpendicular to the plane of incidence ($\alpha \approx 0^\circ$) is confirmed by measurements of several NWs. For $s$-polarization, the photoelectron yield from the top WZ segment is up to $\sim 3$ times larger than that from the ZB segment. We conclude that we can achieve polarization control of the spatial distribution of electron emission using NWs with specific crystal structure and orientation. Both of these parameters can be well controlled for large arrays of nanowires\cite{Anttu14}, making the same concepts scalable beyond single nanowires.

\begin{figure}%
\includegraphics[width=3 in]{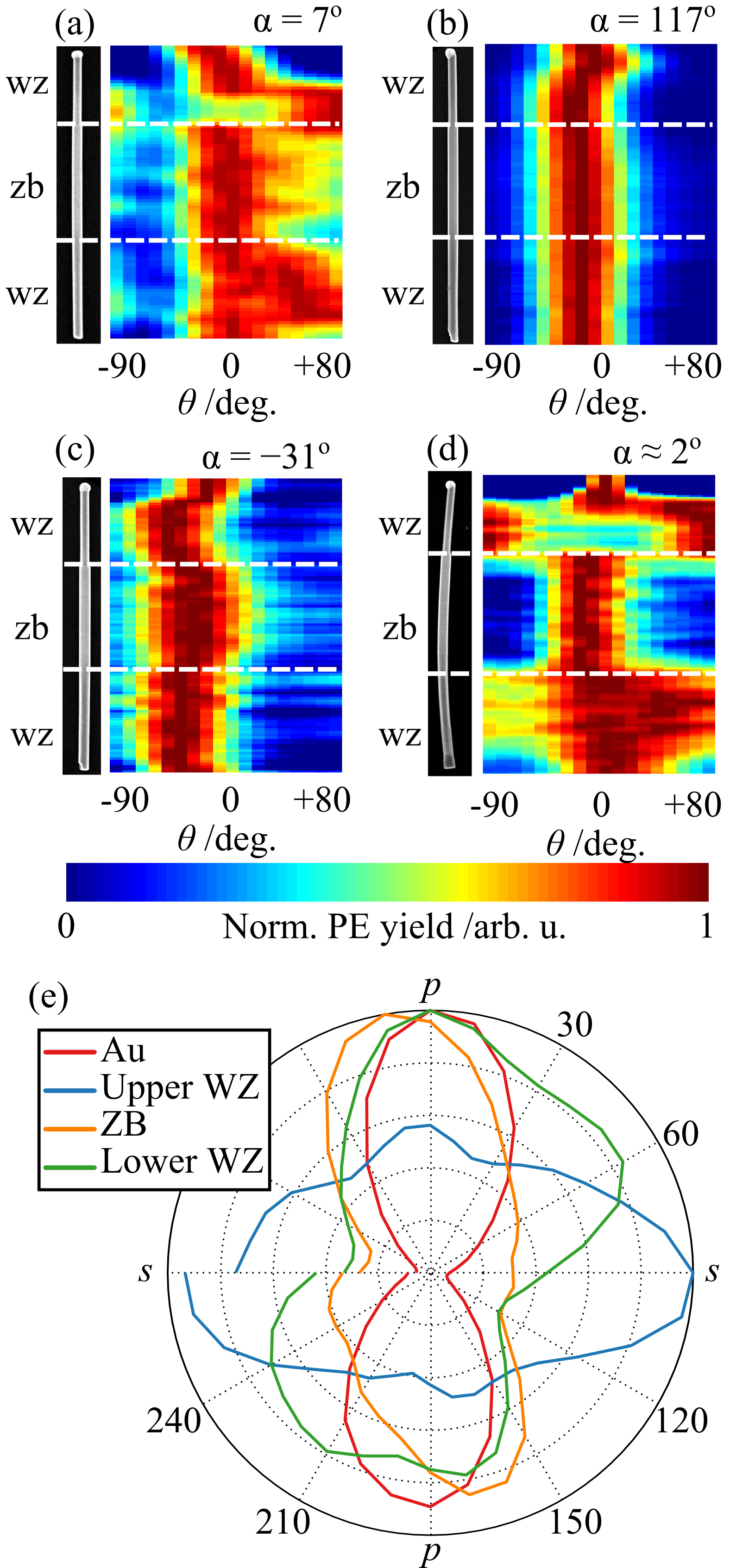}%
\caption{(a--d) Polarization dependence of the photoemission from four different NWs. Each row in the two-dimensional plots shows the normalized photoemission from the corresponding part of the NW as a function of polarization. The left insets show SEM images of the same NWs, with dashed white lines indicating the transitions between different crystal structures. Note that the Au particle appears larger in the PEEM data than in the SEM images due to its much stronger photoemission signal. The rotational angle of each NW is indicated at the top right corner. For the slightly bent NW in (d), the indicated angle $\alpha \approx 2^\circ$ is measured from the top segment. (e) For the NW in (d) we show polar plots of the electron emission from the Au particle as well as the three crystal segments.}%
\label{fig:2Dplots}%
\end{figure}

For a more in-depth analysis of the polarization dependence of the multiphoton photoemission, we acquired series of images where the polarization was rotated in 10$^\circ$ steps (Fig.~\ref{fig:2Dplots}). Two-dimensional plots are displayed, where each row is the normalized photoemission yield at the corresponding position along the NW as a function of polarization angle $\theta$. Dashed white lines are added to indicate the transitions between the different crystal structures, as determined from SEM images of the same NWs (shown to the left of each 2D plot). The 2D plots show how NWs close to perpendicular to the plane of incidence ($\alpha \approx 0^\circ$, a and d) display very different polarization dependences of the emission from different segments. For all NWs, the ZB segment shows maximum emission for a polarization close to $p$. In contrast, the top WZ segment (closest to the Au particle) has its emission maximum for a polarization that strongly depends on the NW orientation and varies from $p$-polarization for NWs close to the plane of incidence (Fig.~\ref{fig:2Dplots}b, $\alpha = 117^\circ$) to $s$-polarization (Fig.~\ref{fig:2Dplots}d, $\alpha \approx 2^\circ$). For all NWs, the bottom WZ segment shows a polarization dependence that resembles a mixture between those of the top two segments. For now, we focus our analysis on the electron emission from the top WZ segment, as this shows the most strikingly different polarization dependence compared to the ZB segment, and address the differences between the two WZ segments later in this letter. In general, the polarizations giving rise to maximum emission in the top WZ segment compared to the ZB segment differ by $\sim 90^\circ$ for NWs perpendicular to the plane of incidence, a difference that rapidly decreases with increasing rotation angle of the NW.

The conclusion from these measurements is that the electron emission from WZ InAs is favored when the electric field is parallel to the long axis of the NW, i.e. along the \textless 0001\textgreater{} crystallographic direction, while the emission from the ZB segment is favored for a polarization close to $p$, regardless of NW orientation. This is especially clear for NWs with $\alpha\approx 0$, for which emission from ZB and WZ are maximized for $p$-polarization and $s$-polarization, respectively. The almost spherical Au particle shows maximum emission for $p$-polarization, regardless of the NW orientation. 

\paragraph{FDTD simulations.} To understand the observed polarization dependences, we have to consider both the nanoscale morphology and the atomic arrangement. Typically, polarization-dependent image contrast in multiphoton PEEM studies has been explained by modeling the electric field concentration from the sample morphology using linear electrodynamics~\cite{Fitzgerald13,Marsell15}. The photoelectron yield has then been assumed to be proportional to the $2n^\mathrm{th}$ power of the instantaneous electric field, with $n$ being the minimum number of photons necessary for photoemission. This method has been successful in explaining the role of nanostructure morphology for the spatial dependence of electron emission. We therefore model the system under study using linear, classical electrodynamics as a first step towards understanding the observed polarization dependence, and then use a microscopic calculation to investigate the role of crystal phase symmetry.

In the case of WZ and ZB InAs, previous theoretical~\cite{De12,Dacal14} and experimental~\cite{Anttu14} studies have shown that for photon energies below 2~eV (as in the present case), there are hardly any differences in the linear optical properties between the three cases of WZ along \textless 0001\textgreater, WZ perpendicular to \textless 0001\textgreater, and ZB. We therefore investigated the linear optical response by assuming a single isotropic dielectric function of InAs, taken from literature data~\cite{Palik85}, and performing finite-difference time-domain (FDTD) simulations of the relevant morphologies. Simulations were performed for $s$- and $p$-polarizations, and the fields for intermediate polarizations were calculated as linear combinations~\cite{Marsell15}. As the photoemission in this case is a 4-photon process, we use the integrated 8$^\mathrm{th}$ power of the electric field norm as a representation of the estimated photoelectron yield. 

\begin{figure}%
\includegraphics[width=3.33 in]{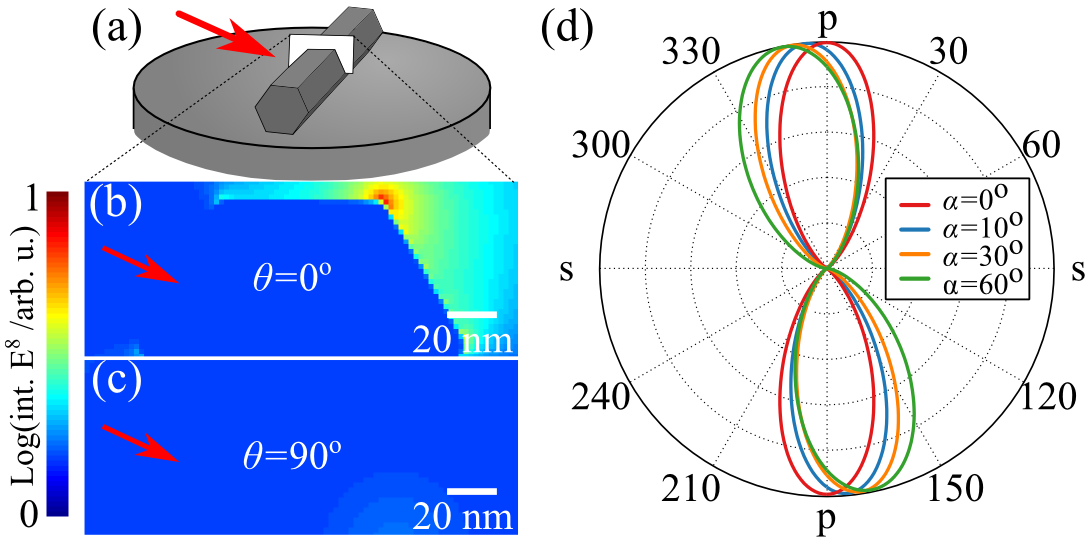}%
\caption{FDTD simulations of InAs nanowires. (a) Schematic illustration of one of the used geometries, where a single InAs NW with a hexagonal cross-section is lying on one of its faces. The time-integrated 8$^\mathrm{th}$ power of the electric field in a cross-section through the plane of incidence is shown on a logarithmic scale in (b--c). (d) Polar plots of the space- and time-integrated 8$^\mathrm{th}$ power of the electric field as a function of polarization angle for three different NWs: the case shown in (a) ($\alpha = 0^\circ$), the same NW rotated 10$^\circ$, 30$^\circ$, and 60$^\circ$, respectively}%
\label{fig:FDTD}%
\end{figure}

Some results of the simulations are shown in Fig.~\ref{fig:FDTD}, illustrating the polarization-dependent field enhancement for nanowires with rotation angles $\alpha$ of 0, 10, 30, and 60$^\circ$. Fig.~\ref{fig:FDTD}b--c  show the integrated 8$^\mathrm{th}$ power of the electric field norm in a cross-section of a NW with $\alpha = 0^\circ$, as indicated in the schematic of Fig.~\ref{fig:FDTD}a. For this case, the field is seen to be concentrated at an edge of the NW for $p$-polarization ($\theta = 0^\circ$), while for $s$-polarization ($\theta = 90^\circ$) the NW barely interacts with the field. The time- and space-integrated 8$^\mathrm{th}$ power of the electric field norm on the upper half of the NW is shown for 4 different rotations as functions of polarization angle in a polar plot. The simulated field is generally enhanced for a polarization close to $p$, regardless of the orientation angle, meaning that the FDTD simulations cannot explain the experimentally observed polarization dependence of the emission from WZ InAs. We also performed FDTD simulations of NWs with different diameters, and of NWs with an edge instead of a face directed towards the PEEM. However, neither of these factors were found to have any major influence on the linear optical response, and in particular none of the simulated geometries could reproduce the maximum around $s$-polarization found experimentally for the photoemission from WZ InAs. We therefore conclude that the polarization-dependent field enhancement around the nanostructure can explain the observed polarization dependence for the Au particle and for ZB InAs, while the emission from WZ InAs is distinctly different. One difference between the FDTD calculations and the experimental studies of the ZB segments is that the contrast between $s$- and $p$-polarization is much weaker in e.g. Fig.~\ref{fig:PEEMimgs}c than in the simulated photoemission intensity shown in Fig.~\ref{fig:FDTD}. However, the calculations essentially assume a morphologically perfect NW. The real NWs have slightly perturbed morphologies and a 1--2~nm surface oxide~\cite{Hjort14}, which has a different structure and can therefore induce a small, presumably isotropic, contribution to the field dependence.

\paragraph{\textit{GW} calculations.} The above approach is based on first calculating the electric field enhancement based on the nanostructure morphology, and then simulating the photoemission assuming an isotropic photoemission process. However, as this model fails to reproduce the experimental results for WZ InAs, we turn to a microscopic description and simulate the polarization-dependent one- two- and three-photon transition rates. 

A one-step treatment of multiphoton photoemission, taking proper account of both intrinsic and extrinsic losses, is currently out of reach for realistic systems due to the inherent complexity of treating continuum and surface states explicitly. Recently, Sirotti {\it et al.}~\cite{Sirotti14} have been able to include surface states in their treatment, but at the same time restricting the discussion to one electron moving in one dimension in a model potential.
However, even using a two-step model of photoemission where the absorption and emission events are considered as independent, any anisotropy in the photoabsorption process should result in a polarization-dependent number of emitted electrons. In the following, we therefore restrict our simulations to a computation of multiphoton photoabsorption rates, i.e. we focus on the polarizarion dependence of the initial stage of the photoemission process.

In order to compute the different multiphoton transition rates, we first need to develop some parts of the numerical machinery. We start the theoretical treatment by considering a one-photon process, where the transition probability from a state with wavevector ${\bf k}$ in valence band $v$ to a state with the same wave vector in conduction band $c$ is given by~\cite{Bassani75}
\begin{eqnarray}
\mathcal{P}_{v{\bf k}\to c{\bf k}} = \frac{2\pi}{\hbar} \left(eE_0\right)^2 \delta_{s_c s_v}\left|\langle \psi_{c{\bf k}}|{\bf e}\cdot {\bf r}|\psi_{v{\bf k}}\rangle\right|^2 \delta(E_c({\bf k}) - E_v({\bf k}) -\hbar\omega).
\end{eqnarray}
We assume the external light field to be of the form ${\bf E}({\bf r},t) = {\bf e} E_0 e^{i\left({\bf q}\cdot {\bf r} -\omega t\right)}$, 
the photon momentum ${\bf q}$ to have a magnitude very close to zero so that only vertical transitions are allowed, and use the length gauge for the particle-field interaction. $\bf e$ is a unit vector defining the polarization of the field, and $E_0$ is the field strength. Defining the transition matrix element ${\bf M}_{cv}({\bf k}) = \langle \psi_{c{\bf k}}|{\bf e}\cdot {\bf r}|\psi_{v{\bf k}}\rangle$ and summing over all bands and $k$-points in the first Brillouin zone, the transition rate per unit time and volume can be written
\begin{align}\label{eq:1ptransition}
W(\omega,{\bf e}) = &\frac{4\pi}{\hbar}\left(eE_0\right)^2 \sum_{cv{\bf k}} \left|{\bf e}\cdot {\bf M}_{cv}({\bf k})\right|^2 \delta(E_c({\bf k}) - E_v({\bf k}) -\hbar\omega). 
\end{align}
The extra factor of two is from a sum over spin, and we have explicitly indicated the dependence of the transition rate on the field polarization ${\bf e}$. For a short pulse with an envelope function $E_0(t)$, we replace the constant $E_0$ in the expression above with the spectral amplitude $E_0(\omega)$.

Extending the above reasoning to multiphoton processes is then possible. At first, we assume that $\omega_1=\omega_2(=\omega_3)=\omega$. Calculations without this restriction are shown in Supporting Information. The resulting formulas for two- and three-photon processes are respectively given by
\begin{align}\label{eq:transitions}
W_{2p}(\omega,{\bf e}) &= \frac{8\pi}{\hbar}\left(eE(\omega)\right)^4 \sum_{c v{\bf k}} \left|\sum_{\alpha}\frac{{\bf e}\cdot {\bf M}_{c\alpha}({\bf k})\, {\bf e}\cdot {\bf M}_{\alpha v}({\bf k})}{E_{\alpha}({\bf k})-E_{v}({\bf k})-\hbar\omega}\right|^2 \delta(E_c({\bf k}) - E_v({\bf k}) -2\hbar\omega) \\
W_{3p}(\omega,{\bf e}) &= \frac{24\pi}{\hbar}\left(eE(\omega)\right)^6 \sum_{c v{\bf k}} \left|\sum_{\alpha\beta}\frac{{\bf e}\cdot {\bf M}_{c\beta}({\bf k})\, {\bf e}\cdot {\bf M}_{\beta\alpha}({\bf k}) \, {\bf e}\cdot {\bf M}_{\alpha v}({\bf k})}{(E_{\beta}({\bf k})-E_{v}({\bf k})-2\hbar\omega)(E_{\alpha}({\bf k})-E_{v}({\bf k})-\hbar\omega)}\right|^2 \nonumber \\
& \quad \times \delta(E_c({\bf k}) - E_v({\bf k}) -3\hbar\omega). \nonumber
\end{align}
The states labeled by $\alpha$ and $\beta$ are intermediate states that have to be summed over in order to include all possible transition pathways, and the prefactors include combinatorial numbers taking into account that the photons are indistinguishable. Finally, the total transition rate is obtained from integrating over frequency, so
\begin{align}\label{eq:transition}
W({\bf e}) = \int d\omega W(\omega,{\bf e}).
\end{align}
In what follows we assume the spectrum of the light pulse to be square shaped, so that $E_0(\omega) = 1$ for $1< \hbar\omega < 2$ eV and zero otherwise.

To apply the above formulas we need to extract the band structure $E_{n}({\bf k})$ and the optical transition matrix elements ${\bf M}_{nm}({\bf k})$ from some microscopic simulation. For this purpose we have performed \textit{GW} calculations using the {\sc abinit}~\cite{Gonze09,Gonze05,Bruneval06} package. Details about the band structure calculations and the Wannier interpolation needed to reach sufficient accuracy at energies significantly above the Fermi level are given in the Methods section and as Supporting Information.

\begin{figure}%
\includegraphics[width=3.33in]{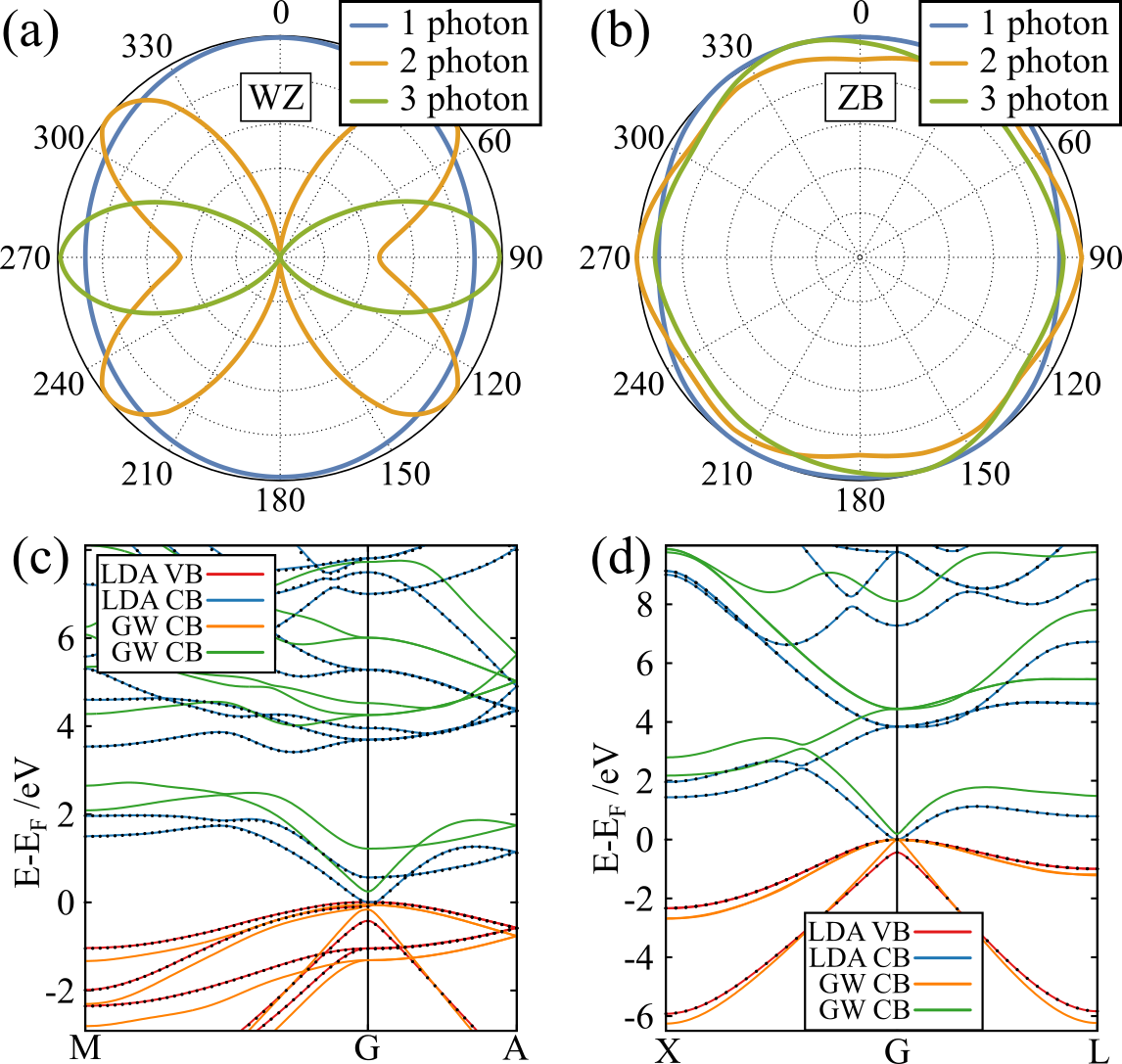}%
\caption{Results of the microscopic calculations. (a--b) Transition rates in bulk WZ (a) and ZB (b) InAs as a function of polarization for one-, two-, and three-photon processes. For the WZ transition rates, a 90$^\circ$ polarization corresponds to the electric field aligned with the c-axis of the crystal, and is therefore equivalent to s-polarization in the experiments for a NW with $\alpha=0^\circ$. (c--d) Band structure of bulk InAs in the WZ (c) and ZB (d) phase. The band structures are calculated explicitly along the indicated $k$-path using DFT within the local density approximation (black dots), and using Wannier interpolation of the bands (red for occupied and blue for unoccupied states). The GW corrected results are shown in orange for occupied states and green for unoccupied states.}
\label{fig:GWfig}%
\end{figure}

The calculated band structures of WZ and ZB InAs are shown in Fig.~\ref{fig:GWfig}c--d. Using these band structures and the associated transition matrix elements, we calculated the polarization-dependent transition rates for one-, two-, and three-photon transitions in WZ and ZB InAs using Eqs.~\ref{eq:1ptransition} and \ref{eq:transitions}. The results for WZ InAs are displayed in Fig.~\ref{fig:GWfig}a, and clearly show that one-photon transitions are isotropic while both two- and three-photon transitions show strong anisotropies, with three-photon transitions being favored for a polarization along the crystal c-axis, which in our experiments corresponds to the NW long axis. In contrast, for ZB InAs (Fig.~\ref{fig:GWfig}b), all transition rates are close to isotropic. The transition rates for WZ InAs were calculated assuming that the c-axis is perpendicular to the plane of incidence. They were also calculated for different field directions within the plane of incidence (i.e., the plane spanned by the crystal vectors $a_1$ and $a_2$), but were found to be independent of the electric field orientation within this plane. For a more detailed discussion of this issue we refer to the Supporting Information.

The key result from these calculations is that while one-photon transitions in this energy interval are close to isotropic in bulk WZ InAs, the multiphoton transitions show strong polarization dependencies. This is in line with previous studies showing that the linear optical properties of WZ InAs are isotropic at photon energies below $\sim 2$ eV. However, our calculations show that the polarization dependence of multiphoton transitions, which have previously not been investigated for WZ InAs, are strongly anisotropic. 

While the multiphoton photoemission from ZB InAs can be fully explained by the NW morphology, multiple mechanisms contribute to the polarization dependence of the electron mission from WZ InAs. First, the electric field can be enhanced at the surface depending on the morphology. For the case of NWs on a substrate, FDTD simulations show that this effect gives the strongest electric fields, and thus the maximum number of excited electrons, for a polarization close to $p$. Furthermore, two- and three-photon transitions from the ground state of WZ InAs show strong, and different, anisotropies as seen from the \textit{GW} calculations (see Fig.~\ref{fig:GWfig}a). Finally, the surface, with possible defects and a thin native oxide layer, further complicates the polarization dependence and can, e.g., introduce a (presumably) isotropic contribution to the transition probability. The combination of these factors explains the range of polarization dependencies seen experimentally for the multiphoton electron emission from WZ InAs. As stated above, for ZB InAs all of the calculated transition rates are close to isotropic, leaving the field enhancement due to sample morphology as the dominating mechanism determininging the polarization dependence of the multiphoton electron excitation rate. 

Finally, we return to the observed differences in the polarization dependence from the top and bottom WZ segments of the NWs (see Fig.~\ref{fig:2Dplots}). Previous studies using low-energy electron diffraction of these NWs~\cite{Hjort14} have shown that the two WZ segments have different surface terminations, due to an overgrowth of the $\left\{1 0 \bar{1} 0\right\}$ facets in the growth process. This difference in surface structure can affect the relative influence of bulk band structure effects and surface effects on the electron emission rate. As the surface contribution will show a different polarization dependence than the bulk contribution, the total polarization dependence can be expected to be a mixture between that of the bulk and that of the surface. The more complicated polarization dependence observed from the lower WZ segment therefore suggests a larger contribution from surface effects to the total electron emission current. In principle, both the morphology and the chemistry of the surface can be engineered, as recently demonstrated~\cite{Hjort14,Knutsson15}.
  
As a final remark, we consider the possibility of induced changes in the optical properties due to the strong excitation. Strong optical excitation of semiconductors has previously been shown to have a prominent effect on the optical properties~\cite{Callan00,Wagner14}. The changes in optical properties have been traced mainly to two origins: heating of the lattice, and state filling in the conduction band. However, lattice heating by electron--phonon interaction takes place on a picosecond time scale. While the excitation, and in some cases the initial decay, of hot electrons can take place on the few-femtosecond timescale, the thermalization of hot electrons responsible for state filling in the conduction band takes $\sim$100 fs, which is much longer than the sub-7~fs pulses used in our study~\cite{Suemoto98}. Furthermore, Callan et al.~\cite{Callan00} found only small optically induced changes to the dielectric function of GaAs as long as the excitation fluence was below 50~mJ~cm$^{-2}$, which is a factor of $\sim$50 higher than what was used in our experiments. Changes in the optical properties induced by high excitation densities are therefore not believed to affect the photoemission intensity in our experiments.

In summary, we have studied the polarization-dependent multiphoton electron excitation in wurtzite (WZ) and zincblende (ZB) InAs. Exciting InAs NWs with few-cycle near-infrared pulses and detecting the emitted electrons from WZ and ZB segments with $\sim 50$~nm spatial resolution, we found that the electron yield from the ZB segment of NWs perpendicular to the plane of incidence was maximized for $p$-polarization. This can be attributed to the field enhancement at the NW surface, as confirmed by FDTD simulations. For WZ segments of the same InAs NWs, the electron yield showed a strikingly different polarization dependence. This effect cannot be explained by linear electrodynamics; however, we calculated multiphoton transition rates based on self-consistent quasi-particle \textit{GW} band structure calculations. These calculations show strong polarization anisotropies in the two- and three-photon transition rates in bulk WZ InAs, while the linear interaction is close to isotropic. The anisotropy of the multiphoton absorption present in WZ InAs but not ZB InAs thus explains the experimentally observed differences. The need for microscopic simulations of nonlinear light--matter interactions is in stark contrast to multiphoton photoemission studies on metals and semiconductors with cubic symmetry, where linear classical electrodynamics simulations in combination with an isotropic photoemission process have typically been able to reproduce the experiments. 

Our study shows how crystal structure can be added to the set of methods to control multiphoton electron excitations in matter. This is of particular interest for III--V NWs, as they can be made with a high level of control of crystal structure, surface chemistry, and morphology. This opens up for applications where the excitation of hot electrons can be controlled by a range of parameters. These electrons can for example be used to control electron transport across a semiconductor--dielectric interface, or to control chemical reactions on the surface\cite{Mukherjee13}. In total, this could open up new pathways in hot electron-based devices for optoelectronics and catalysis.

\section*{Methods}
\paragraph{PEEM experiments.} The experiments were performed at pressures below $1\cdot 10^{-8}$ mbar using a commercial PEEM instrument (Focus GmbH) routinely achieving sub-50~nm spatial resolution. The instrument is equipped with a linear PEEM column with 3 electrostatic lenses, as well as a retarding field energy analyzer and a detector consisting of a multi-channel plate, a fluorescent screen, and a peltier-cooled CCD camera operating at $-20\;^\circ$C. All PEEM images presented were acquired using an extractor voltage of 12.5 kV and in energy-integrating mode, with typical acquisition times of 60 seconds per frame. The light source was an optical parametric chirped pulse amplification (OPCPA) system delivering sub-7~fs pulses at a repetition rate of 200~kHz with a center wavelength of~850 nm. The OPCPA system was similar to what has been described previously~\cite{Miranda12}. A broadband Ti:Sapphire oscillator is used to both provide seed pulses to the OPCPA and to optically synchronize an Yb-fiber laser chain. The frequency-doubled output from the fiber amplifiers then acted as the pump for two non-collinear optical parametric amplification (NOPA) stages. For additional control of the output, a pulse shaper was built and used to shape the phase of the oscillator pulse seeding the NOPA stages. The pulse shaper consisted of a pair of gratings, two collimating mirrors, and a one-dimensional LCD-based spatial light modulator with which the phase of each spectral component could be controlled. Dispersion control of the output pulse was realized using double chirped mirrors and a pair of glass wedges. The resulting pulse had a duration down to 6.1~fs, with two minor (\textless 10\% of the peak intensity) satellite pulses approximately 10 and 20~fs from the main pulse, as retrieved from a d-scan measurement~\cite{Miranda12} and shown in Fig.~\ref{fig:Schematic}b--c. The polarization was controlled using a broadband half-wave plate, and the laser beam was focused by a 20~cm focal length achromat lens to a spot size of approximately $50\times 100 \;\upmu \mathrm m^2$ on the sample. Amplified pulses were needed for the experiments due to the low field enhancement expected from semiconductor nanostructures. Compared to previous PEEM studies of metallic nanostructures using pulses with around 800~nm wavelength, the average power of $\sim 10$~mW used in our study is similar. However, the short pulse duration and repetition rate of 200~kHz gives a peak intensity of $\sim 10^{11}$~W~cm$^{-2}$, which is one to two orders of magnitude higher than what has been reported for multiphoton PEEM on metallic structures~\cite{Schertz12,Sun13,Marsell15}. The incidence angle of the laser beam was 65$^\circ$ to the sample normal, as shown in the schematic of Fig.~\ref{fig:Schematic}a.

\paragraph{Sample preparation.} Nanowires were grown on InAs substrates by metal-organic vapor-phase epitaxy using the method described by Lehmann \textit{et al.}~\cite{Lehmann13}, with which the crystal structure can be controlled by adjusting the group V flow. For optimum homogeneity of the nanowires, they were grown in an ordered array defined by electron beam lithography. A convoluted dark-field TEM image of a typical NW is shown in Fig.~\ref{fig:Schematic}d. The dark-field imaging clearly shows the abrupt interfaces between the ZB and WZ crystal structures, as well as the twinning within the ZB segment. The nanowires were transferred from the growth substrate onto a piece of highly doped, native oxide-covered Si(001) wafer by mechanical break-off. The same nanowires imaged in the PEEM experiments could afterwards be identified in an SEM (Hitachi SU8010) for determination of the morphology of each individual nanowire.

\paragraph{Calculations.} Two types of microscopic calculations were performed using the \textsc{abinit} package: The first uses DFT~\cite{Hohenberg64,Kohn65} in the local density approximation (LDA), which is numerically fast, but neglects non-local electronic correlations. The second approach includes many body effects by first using the HSE hybrid functional to correct the band gap and then performing a self-consistent quasi-particle $GW$ calculation. Due to the high computational demand of these calculations and the need for an exceptionally dense grid of $k$-points, direct $GW$ calculations are practically impossible. We have therefore used the post-processing package {\sc wannier90} to interpolate the matrix elements, starting from a $k$-point grid of relatively low density. This requires the construction of Wannier functions from the Bloch functions used by {\sc abinit}. To verify that the Wannier functions are constructed properly and are well converged, we calculated the band structure of InAs in both the zincblende and wurtzite structures (see Supporting Information). As an additional test, we also computed the dielectric function of InAs both using {\sc abinit} and with the help of {\sc wannier90}, obtaining excellent agreement (not shown).

Classical electrodynamics simulations were performed using the FDTD Solutions software from Lumerical. The simulations took into account the substrate. The calculations were performed with a broadband modulated Gaussian total field scattered field plane wave source. The excitation spectral range was set to match the spectral range of the laser used in the experiment. The simulation space was closed with Perfectly Matched Layers.

\section*{Notes}
The authors declare no competing financial interest.

\begin{acknowledgement}

This work was supported by the Swedish Research Council (VR), the Swedish Foundation for Strategic Research (SSF), the Crafoord Foundation, the Knut and Alice Wallenberg Foundation, and the European Research Council (ERC) grants ElectronOpera and PALP.

\end{acknowledgement}

\begin{suppinfo}

Interferometric time-resolved PEEM measurements, and detailed descriptions of the band structure calculations and the use of Wannier interpolation are available as Supporting Information.

\end{suppinfo}


\bibliography{InAs}

\end{document}